# Interferometer measurements in interstellar communications:
# methods and observations

*William J. Crilly Jr.*

Green Bank Observatory, West Virginia, USA

*Abstract*— Extraterrestrial communication signals are hypothesized to be present in an extensive search space. Using principles of communication theory and system design, methods are studied and implemented to reduce the signal search space, while considering intentional transmitter detectability. The design and observational work reported in this paper adds material to previous related reports: arXiv:2105.03727v2, arXiv:2106.10168v1, arXiv:2202.12791v1, and arXiv:2203.10065v1. In the current work, a two-element radio interferometer telescope and receiver algorithms are utilized to perform differential angle-of-arrival and multi-bandwidth measurements of $\Delta t\ \Delta f$ polarized pulse pairs. The system enhances extraterrestrial signal detectability, while reducing signal false positives caused by noise and radio frequency interference. Statistical analysis utilizes a Right Ascension filter spanning celestial coordinate ranges that include the previously determined anomalous celestial direction: $5.25 \pm 0.15$ hr Right Ascension, $-7.6° \pm 1°$ Declination. Observations were conducted during a duration of 61 days, comprising 244 hours of interferometer measurements.

## I. Introduction

In the Search for Extraterrestrial Intelligence (SETI), many hypotheses and associated experimental methods have been proposed to analyze and try to explain the detection of unknown-cause apparent extraterrestrial radio signals, in the presence of noise, Radio Frequency Interference (RFI) and natural astronomical objects.[1][2][3] Interferometers have been used in SETI since 1975.[4]

In the experimental work reported in this paper, a two-element interferometer and signal processing system is used to implement an automated RFI excision method, narrow-bandwidth multi-pulse detection algorithm, and a differential phase measurement of two-tone pulsed narrowband signals, facilitating a differential celestial angle-of-arrival measurement method.

Previous experimental work has suggested an anomalous celestial pointing direction, having an apparently significant number of $\Delta t\ \Delta f$ polarized pulse pairs, at $5.25 \pm 0.15$ hr Right Ascension (*RA*), $-7.6° \pm 1°$ Declination (*DEC*).[5][6][7][8]

The current work reported here presents measurements that tend to confirm the presence of anomalies in the same celestial pointing direction as the previously determined anomalous direction.

Many explanatory hypotheses exist and are the topic of ongoing and/or further work.

## II. Hypothesis

The hypothesis in this work is similar to those in previous work, [5][6][7][8], with the primary difference that a radio interferometer is used to filter narrow bandwidth received signals using differential RF phase measurements across interferometer elements. The phase measurements identify differential angles-of-arrival of spectral components of discoverable communication-like signals that are expected to be rarely found in random noise.

*Hypothesis:* Narrow-bandwidth energy-efficient interstellar communication signals are expected to be explained using an Additive White Gaussian Noise (AWGN) model, while using a radio interferometer with receiver algorithms that reduce RFI, and provide Signal-to-Noise Ratio (SNR) reduction. Receiver filters provide differential angle-of-arrival measurements of $\Delta t\ \Delta f$ polarized pulse pairs having celestial coordinates centered on a prior anomalous celestial direction: $5.25 \pm 0.15$ hr *RA* and $-7.6° \pm 1°$ *DEC*. In the current work, the time difference $\Delta t$, between narrow bandwidth pulses in a pair, is equal to zero seconds.

Discoverable communication-like $\Delta t\ \Delta f$ polarized pulse pairs observed with the interferometer are expected to not indicate measurements similar to those expected from an AWGN signal model. Falsification of the stated AWGN-cause hypothesis therefore compels activities that produce alternate and auxiliary hypotheses, topics of ongoing and/or further work.

Salient conjectures in this hypothesis and experimental design are that interstellar communication signals
(1) have high information capacity [5],
(2) are energy efficient [5],
(3) occupy a wide bandwidth [5],
(4) indicate measurements approaching those of AWGN,[5][9]
(5) exceptionally transmit high SNR narrow bandwidth pulse pair components, readily discoverable in random noise, [5][10],
(6) minimally interfere with known and likely communication systems[5], and
(7) indicate repetition at celestial pointing directions during long duration experiments[7].

Details of transmitter design rationale are described at III. Transmitter Design in [5].

---

William J. (Skip) Crilly Jr. is a Volunteer Science Ambassador in Education & Public Outreach of the Green Bank Observatory. email: wcrilly@nrao.edu

## III. METHOD OF MEASUREMENT

### Objectives

*Facilitate hypothesis falsification*

Given an interest to attempt to falsify the hypothesis [11], the experiment is designed to attempt to remove all contributions to the detection of non-AWGN signals. Following the experiment, if $\Delta t$ $\Delta f$ polarized pulse pairs appear anomalous with statistical power, then the hypothesis is falsified, compelling alternate and auxiliary hypotheses to be proposed and tested.

*Right Ascension filter*

The AWGN-explanation hypothesis is falsified when anomalies are discovered at particular values of pointing *RA*, after relatively long duration experiments.

*Machine design*

Measurement machines ideally should be automated, from antenna voltage measurement, to the presentation of results in figures. Traceability, repeatability and equipment and human error tracking can then proceed. For example, RFI confounds almost all hypotheses related to SETI. Consequently, the machine that excises suspected RFI should include a repeatable and documented process, and include a process that stores RFI that has been excised, to facilitate follow-up and future RFI hypothesis testing.

*Associated measurements*

Various measurements may aid hypothesis falsification. Examples of associated measurements include RF frequency, frequency separation between simultaneous narrowband pulses, SNR, power measurements in various bandwidths, interferometer complex cross-correlation, and RFI excision measurements.

### Measurement System

*Antenna system*

The antenna system is located in a rural location having low and controlled local RFI.

The two interferometer antenna elements are offset-fed paraboloids, each approximately eight-feet in diameter, separated along an East-West baseline at a distance of 32 wavelengths of 1425 MHz. Each antenna feed comprises a pair of two half-wave spaced, phased folded dipoles per linear polarization, combined in a 90 degree hybrid combiner, to generate right-hand and left-hand circular polarized voltages per feed. The four dipoles are placed in a truncated conical shaped reflector. Low noise amplifiers are located at each feed. In the current work, the element antennas' right-hand circular polarized signals are processed.

Interferometer element performance is tested using a combination of artificial sources driving helical test antennas, and astronomical objects. Each element's Azimuth and Elevation is measured and transmitted to instrumentation. Antenna elements are set to a *DEC* of -8°, and Azimuth at 180°, confirmed with continuum and interferometer correlation measurements of astronomical object NRAO 5690.

*Receivers*

Right-hand circular-polarized voltages from the feed are applied to downconverters that output in-phase and quadrature-phase (I+Q) zero-IF baseband signals, using a 1425 MHz local oscillator, phase locked to an oven-controlled crystal oscillator, stabilized to GPS satellite signals. The two I+Q signals from the two interferometer elements are digitized in a four channel synchronous 8-bit digitizer, triggered by a three second interval pulse, synchronized to GPS satellite signals. UTC time is produced by a GPS receiver output, in a coded signal centered at one kHz, and applied at baseband to one channel of the four channel digitizer, to ensure the sample time integrity of the antenna signals.

Four complex FFTs are calculated per 3 s interval, two adjacent-time, one for each antenna element, at a bin bandwidth of 3.7Hz and 0.27s interval per FFT. Measurements of each candidate are made, e.g. MJD, *RA*, SNR, RF Frequency, $Log_{10} \Delta f$ /MHz, RFI excision values.

*RFI excision*

First-level filtered candidate signals cover the frequency range of 1398 to 1451 MHz, with RFI excision implemented at the time of data capture, and described as follows.

Due to the regulatory spectral allocation of human communication systems, RFI is often observed to be concentrated in RF frequency. A hypothetical high capacity interstellar communication system, on the other hand, might be expected to have wide-bandwidth noise-like characteristics. These differences have led to the design of an excision method that stores measurements of a limited number of repetitive narrow bandwidth received signals in a file separate from the file that stores desired candidate signals.

The FFT receiver is partitioned to create 256 FFT bin segments, to determine the presence of suspected RFI. As a four hour duration file is filled with candidate measurement records, the number of candidate signals is counted in each segment. A per segment record count threshold, set to a value of ten, is implemented. The segment count threshold sets the count at which the segment is determined to be an RFI corrupted segment. Subsequent records in this segment are saved to an RFI file, in lieu of a first-level candidate file, up to two hundred RFI records per segment.

Each measurement record in the candidate file contains segment count values that estimate the segment spacing from the measured pulse candidate segment to the nearest excised RFI segment, above and below the candidate segment.

The threshold to excise a segment is set to balance the excision of potentially desired segments with the need to excise suspected RFI segments. When RFI is present, RFI excision of RFI segments typically stabilizes within a few minutes of the beginning of a four hour file being created. An RFI maximum count per 0.27 s, set to one hundred, prevents files from becoming too large with sustained RFI.

RFI is sometimes observed at, or near, harmonics of 1 MHz and 100 kHz. Consequently, a fixed filter is utilized that rejects candidates within 15 kHz of harmonics of 100 kHz.

In a zero-IF receiver, RFI is sometimes observed near the region close to the LO frequency. The range of 1424 to 1426 MHz is therefore excised.

Second-level filtered candidate signals have low and high end-of-range excision, resulting in the overall frequency range of 1405 to 1435 MHz. This second-level excision was implemented as a result of intermittent RFI discovered near the start and end of the first-level frequency range, after the 61 day experiment began.

### Interferometer phase measurement

RF amplitude and phase is calculated in each 3.7 Hz bandwidth FFT bin of each interferometer element. The difference in RF phase measurement between the East and West interferometer elements is then calculated, represented by $\Delta\Phi_{INT.RF}$. The absolute value of the difference between $\Delta\Phi_{INT.RF}$ values of two $\Delta t=0$ $\Delta f$ separated components of polarized pulse pairs is calculated, expressed as $|\Delta\Delta\Phi_{INT.RF}|$.

$|\Delta\Delta\Phi_{INT.RF}|$ is further described in **First and second-level candidate filters**, below.

### Interferometer complex correlation

The use of a high frequency resolution complex FFT in each interferometer element receiver becomes part of an FX type correlator, i.e. a cascade of an FFT per element receiver, a complex conjugate multiply per frequency bin and an IFFT to calculate the correlation per delay tap. [13] Delay taps have an increment of 16 ns.

The complex correlation provides a sensitive measurement of wide bandwidth emission from natural celestial sources. Natural celestial sources potentially transmit narrow bandwidth spectral components that might appear similar to intentional communication signals.

### Multi-bandwidth power measurements

Average power measurements are performed in 954 Hz and 50 MHz bandwidths. The 954 Hz bandwidth power measurement provides the noise value, after bandwidth scaling, in the measurement of 3.7 Hz bandwidth SNR.

### First and second-level candidate filters

First-level candidate records are defined to be a set of measurements that pass the first-level filter, i.e. FFT output bins that measure SNR greater than 8.5 dB on each interferometer element, and absence of RFI excision. The measurements of first-level candidates are saved to a four hour duration file at the time of signal acquisition. First-level candidates are produced at a rate of approximately fifteen records per 3 seconds.

In second-level candidate filtering, a first-level candidate input file is chosen, per MJD day, to cover the *RA* direction of interest, 5.25 hr *RA*. Second-level candidate output records are defined to be a set of records culled from first candidate records that meet the additional requirements of RF frequency range, pulse pair frequency spacing $\Delta f$ <100 kHz, and interferometer $|\Delta\Delta\Phi_{INT.RF}|$ < 0.1 radian. The number of second-level output candidates was 1,280 during the 61 day experiment. Second-level output candidates' measurement records are sorted by increasing $|\Delta\Delta\Phi_{INT.RF}|$, to facilitate binomial-model likelihood calculations.

$|\Delta\Delta\Phi_{INT.RF}|$ is defined as follows,

$$\Delta\Delta\Phi_{INT.RF}(i) = (\Phi_{INT.RF.WEST}(i) - \Phi_{INT.RF.EAST}(i))$$
$$-(\Phi_{INT.RF.WEST}(i-1) - \Phi_{INT.RF.EAST}(i-1)), \quad (1)$$

where $\Phi_{INT.RF}(i)$ is the interferometer RF phase measurement, per FFT bin, per element. The count index i increments with the FFT bin-sorted SNR East > 8.5 dB and SNR West > 8.5 dB $\Delta t=0$ polarized pulse pair candidates.

### SNR threshold

The SNR threshold of 8.5 dB is approximately 3 dB below the SNR threshold used in previous work [5][6][7][8]. The difference is explained as follows.

Assuming a Rayleigh-distributed amplitude probability per FFT bin, given a single receiving antenna element, the following amplitude density applies [14],

$$p_{RA}(r) = (r/\sigma^2) \exp(-r^2/2\sigma^2), \quad (2)$$

with r ≥ 0, equal to the Gaussian noise amplitude density variable and $\sigma^2$ the Gaussian noise variance. The cumulative probability of a pulse event having amplitude from r to infinity is equal to

$$P_{RA}(r) = \exp(-r^2/2\sigma^2). \quad (3)$$

In a two element interferometer with an FFT having $N_{FFT}$ frequency bins, after RFI excision, the number of first-level candidates having a cumulative amplitude in both elements above the threshold r, per FFT interval, is

$$N_{CANDIDATES} = N_{FFT} \exp(-r^2/2\sigma^2) \exp(-r^2/2\sigma^2), \quad (4)$$

resulting in a factor of two in the exponential,

$$N_{CANDIDATES} = N_{FFT} \exp(-2r^2/2\sigma^2). \quad (5)$$

Given a signal processing goal to produce a fixed number of potential signal candidates per FFT calculation, i.e. $N_{CANDIDATES}$ (5), the SNR threshold, $r^2/2\sigma^2$, may be reduced by 3 dB, compared to using a single antenna and receiver. A planned third interferometer element is expected to provide an additional 1.7 dB reduction in candidate SNR threshold.

## IV. OBSERVATIONS

Observations were conducted between Dec. 6, 2023 and Feb. 4, 2024. The signal processing algorithms provided a total of 1,280 filtered and sorted measurement records, plotted in **Figs. 1-13**. The text below each of **Figs. 1-13** describe measurements of the 1,280 candidate pulse pairs, as they relate to the previously stated **Hypothesis** of this work. All figure image files were produced in the machine second-level candidate signal processing step, without human intervention.

In general, **Figs. 1-4** present measurements that tend to provide evidence to falsify the AWGN hypothesis, while **Figs. 5-13** present measurements that may be used to seek alternate and auxiliary explanations related to the measurements presented in **Figs. 1-4**.

**Fig. 1** indicates the count per *RA* bin of filtered and sorted $\Delta t=0$ $\Delta f$ polarized pulse pairs produced by the signal processing filters described in **METHOD OF MEASUREMENT**.

**Fig. 2** indicates the statistical power of sorted $\Delta t=0$ $\Delta f$ polarized pulse pairs anomalies across *RA* bins, observed in **Fig. 1**.

**Fig. 3** indicates the $Log_{10}$ $\Delta f$ /MHz per *RA* bin, to seek anomalies, in attempts to falsify the Poisson distribution expected in an AWGN explanatory hypothesis. The anomalies of low $\Delta f$ values in *RA* bins 51 and 52 indicate that these *RA* bin responses contain signals dissimilar from AWGN.

**Fig. 4** indicates the degree to which the cumulative effect of many low-valued $\Delta f$ pulse pairs are presented in each *RA* bin, by calculating the sum of $\text{Log}_{10} \Delta f$ /MHz of pulse pairs observed per *RA* bin. *RA* bins 51 and 52 indicate that pulse pairs exist that are not explained by Poisson-distributed $\Delta f$ frequency spacing, in AWGN. The calculation of $\Delta f$ Likelihood in AWGN is presented in [5] Appendix C.

**Fig. 5** indicates the East interferometer element relative average power measured at the same time in the 954 Hz wide segment that spans the FFT bin of the associated sorted $\Delta t=0$ $\Delta f$ polarized pulse pair. Anomalous 954 Hz bandwidth power may present ways to explain various causes for outlier pulse pairs. Power is integrated over 0.27s.

**Fig. 6** has a measurement rationale similar to that in **Fig. 5**. A wider bandwidth measurement of the East interferometer element relative average power, at 50 MHz, helps identify various explanatory causes of anomalous sorted $\Delta t=0$ $\Delta f$ polarized pulse pairs. Power averaging time is 0.27 s.

**Fig. 7** indicates the MJD time that the $\Delta t=0$ $\Delta f$ pulse pair candidate was measured.

**Fig. 8** indicates the RF frequency of the higher frequency component in the associated $\Delta t=0$ $\Delta f$ polarized pulse pair.

**Fig. 9** indicates the binomial model event probability that a pulse pair event will randomly be present in an *RA* bin, given the non-uniform sampling of *RA* in the 61 day experiment.

**Fig. 10** indicates the $\text{Log}_{10}$ Likelihood of SNR per pulse pair, per *RA* bin. The calculation of SNR Likelihood in AWGN is presented in [5] Appendix B.

**Fig. 11.** indicates the $|\Delta\Delta\Phi_{INT.RF}|$ measurement values per pulse pair in each *RA* bin.

**Fig. 12** and **Fig. 13** indicate the number of 954 Hz segments between the nearest identified RFI segments and the segment containing the RF frequency of the $\Delta t=0$ $\Delta f$ polarized pulse pair.

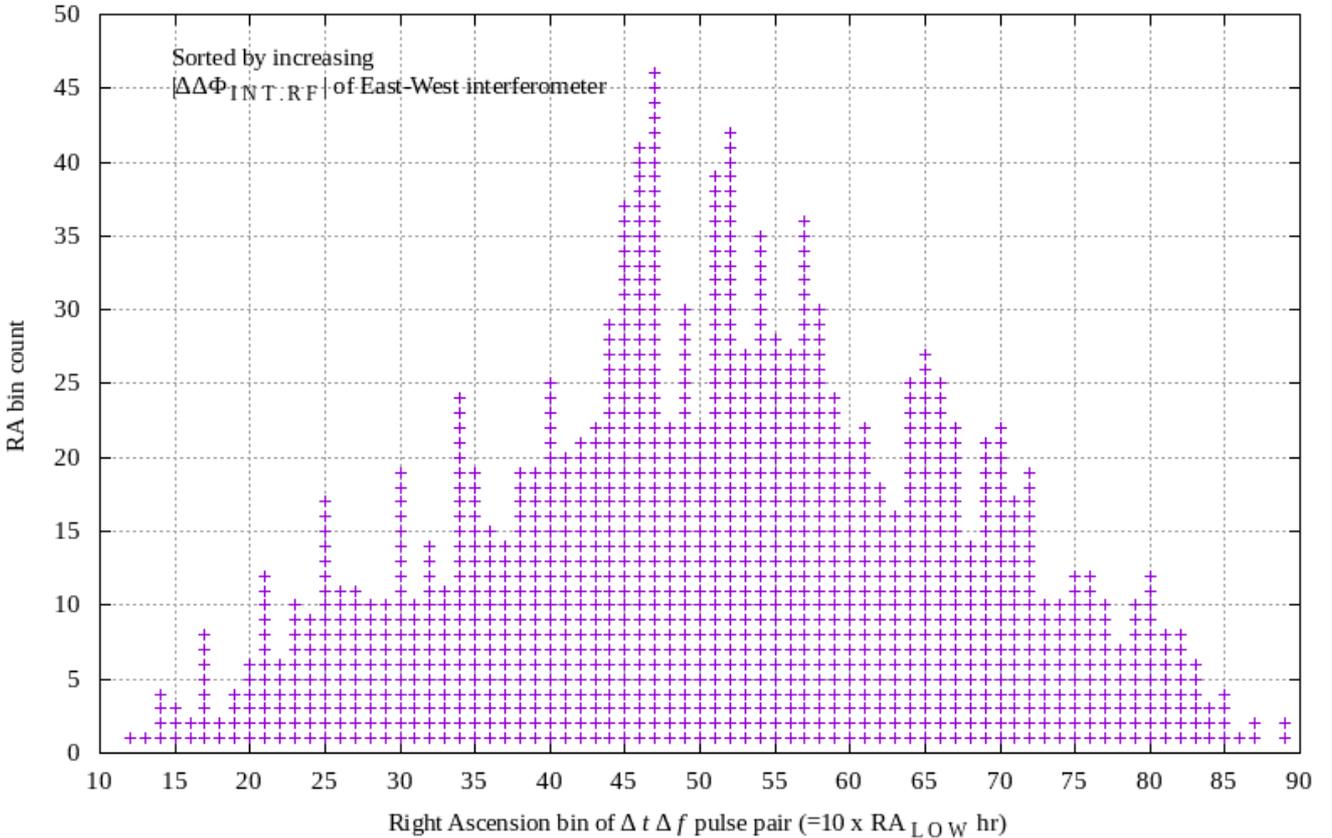

**Figure 1:** The count of 3.7 Hz bandwidth filtered $\Delta t$ $\Delta f$ polarized pulse pairs, having $\Delta t=0$, $\Delta f < 100$ kHz and pulse pair absolute $\Delta f$ differential interferometer phase, $|\Delta\Delta\Phi_{INT.RF}| < 0.1$ rad., is plotted per *RA* bin. The experiment's duration was 244 hours during 61 days. The 5.15 and 5.25 hr *RA* bins, bins 51 and 52, indicate apparent anomalous counts of $\Delta t=0$ $\Delta f$ pulse pairs. 5.25 ± 0.15 hr *RA* is the previously determined *RA* direction of interest. The $|\Delta\Delta\Phi_{INT.RF}| = 0.1$ rad. value corresponds to a sky pointing angle difference, of $\Delta f < 100$ kHz pulse pairs, of 0.028°. Anomalies are also apparent in bins 45, 46 and 47. The statistical power of count anomalies is calculated and shown in **Fig. 2**.

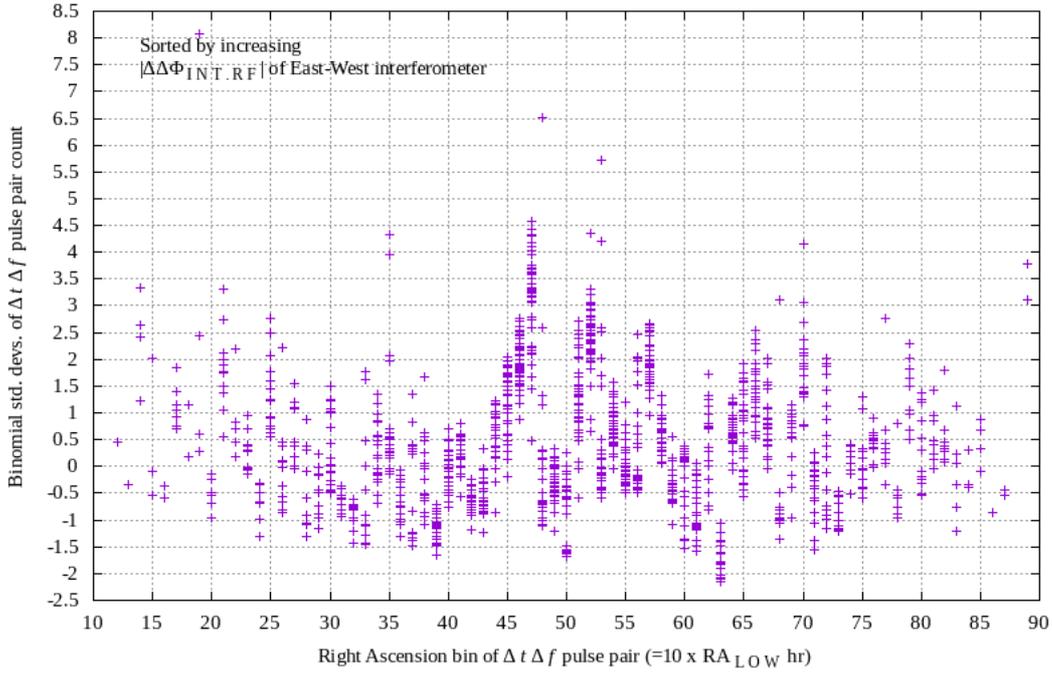

Figure 2: $\Delta t \Delta f$ pulse pair-associated Binomial std. devs. per RA bin vs. RA bin ($\Delta RA$ bin = 0.1 hr.);
# of points = 1280 ; MJD range = 60284.179902 - 60344.1642766 ;
$Log_{10} \Delta f/MHz$ Max = -1.000 ; $|\Delta\Delta\Phi_{INT.RF}|$ Max = 0.100 radians

**Figure 2:** Each measured count of $\Delta t=0$ $\Delta f$ polarized pulse pairs, described in **Fig. 1**, has an associated binomial-distributed Likelihood, given the number of increasing $|\Delta\Delta\Phi_{INT.RF}|$ sorted trials, measured count per *RA* bin and event probability. An expected value of count mean and standard deviation is calculated at each trial, for each *RA* bin, assuming an AWGN-explanatory cause and binomial statistics. These AWGN model statistics, together with measured event counts, are used to determine an effect size, based on the method in Cohen's **d** = $\Delta$mean / std.dev. The dimensionless method indicates the statistical power of anomalous counts in *RA* population bins. A Cohen's **d** value of 0.8 is considered to be a "large" effect size.[12] *RA* bins 51 and 52 indicate Cohen's **d** values surrounding 1 and 2.5 respectively.

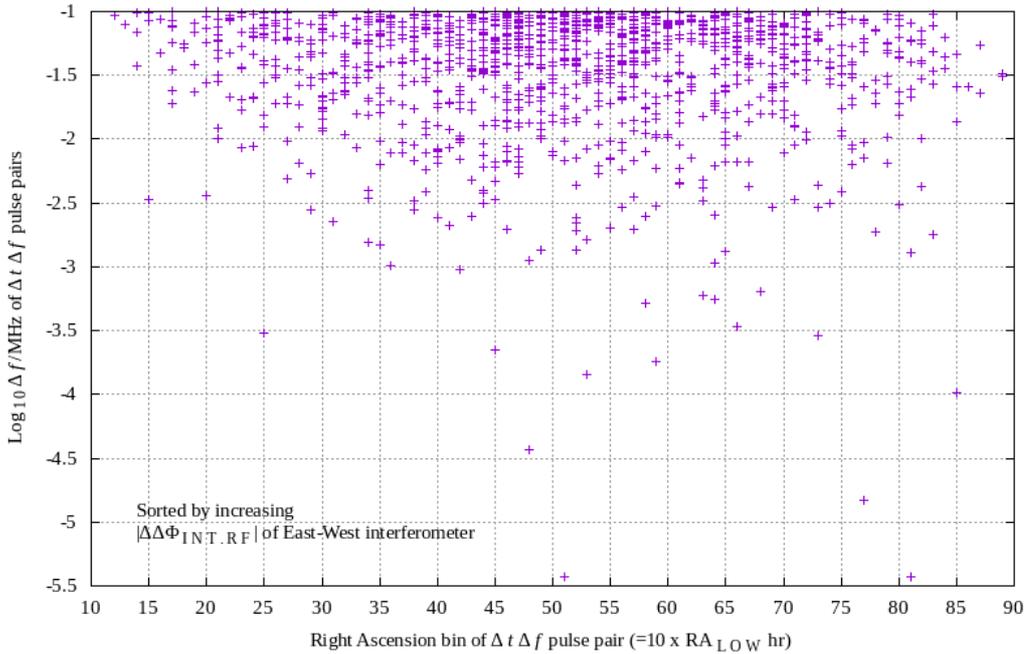

Figure 3: $\Delta t \Delta f$ pulse pair-associated $Log_{10} \Delta f/MHz$ vs. RA bin ($\Delta RA$ bin = 0.1 hr.);
# of points = 1280 ; MJD range = 60284.179902 - 60344.1642766 ;
$Log_{10} \Delta f/MHz$ Max = -1.000 ; $|\Delta\Delta\Phi_{INT.RF}|$ Max = 0.100 radians

**Figure 3:** $Log_{10} \Delta f$ /MHz indicates anomalous close-spaced frequency components (one pulse pair having $\Delta f$ =3.7 Hz, four pulse pairs having $\Delta f$ < 2.5 kHz), in *RA* bins 51 and 52, respectively. The binomial distribution Likelihood of a noise-cause calculates to *pr.*=0.047 (1 seen in 2 trials at event *pr.* 0.024), and *pr.*=0.011, (4 seen in 38 trials at event *pr.* 0.024) respectively.

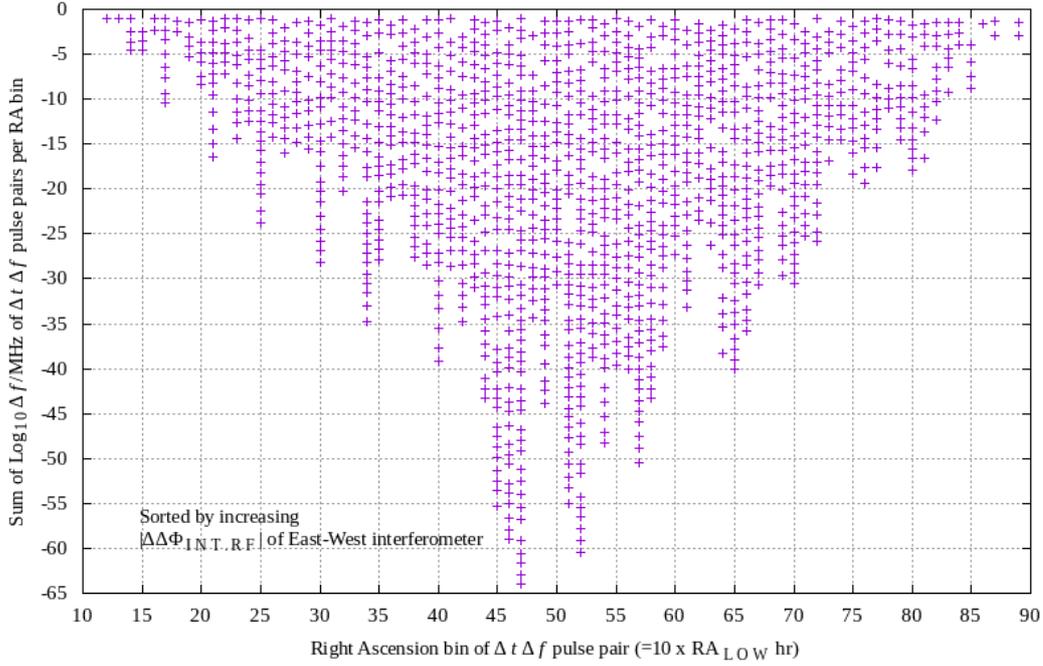

**Figure 4:** The sum of $Log_{10}$ $\Delta f$ /MHz per *RA* bin indicates the repetitive significance of low-valued $\Delta t=0$ $\Delta f$ <100 kHz pulse pairs. *RA* bins 51 and 52 indicate anomalies. Anomalies in other *RA* bins is a topic of further experimental work. The *DEC* beamwidth of each interferometer element is approximately 8º, greater than the *DEC* beamwidths used in previous experiments (≈0.6º-3º0),[5][6][7][8], suggesting a possibility that a similar explanatory cause might be present across *RA* bins, i.e. at different celestial coordinates.

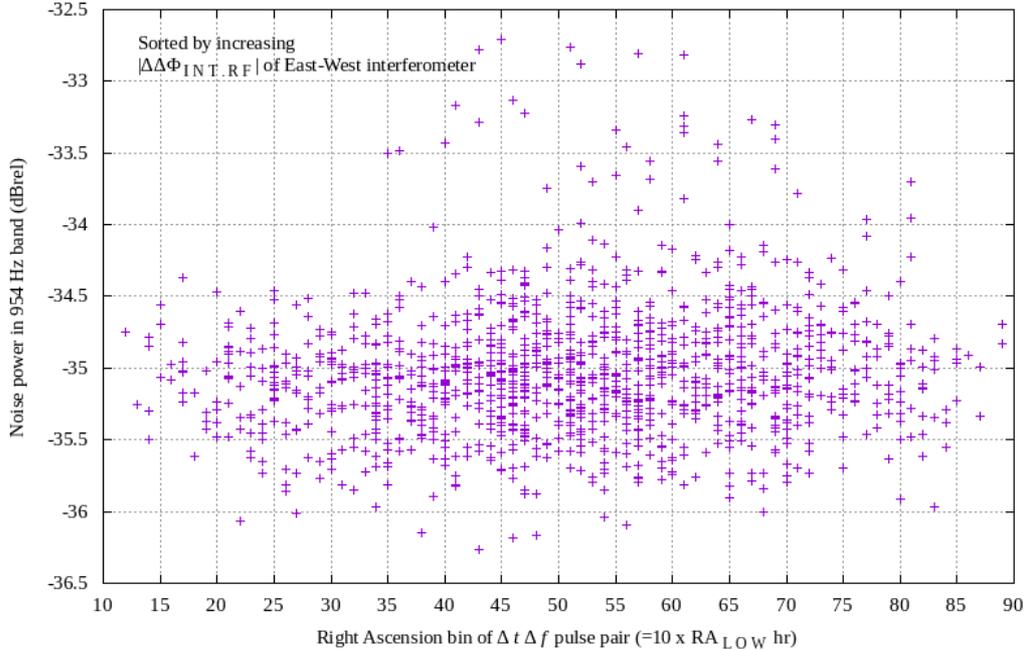

**Figure 5:** 954 Hz bandwidth integrated East element power is measured, during the 0.27 s FFT time window of the pulse pair, in the 256 bin FFT spectral segment containing the anomalous $\Delta t=0$ $\Delta f$ polarized pulse pair. The 954 Hz bandwidth power measurements help determine if 954 Hz bandwidth power levels might explain narrow bandwidth spectral outliers, observed in **Fig. 1-4**. Approximately twenty anomalous *RA* bin 51 and 52 pairs are observed in **Fig. 1**, while approximately seven anomalous 954 Hz bandwidth *RA* bin 51 and 52 pairs are observed in **Fig. 5**. Work is underway to develop explanatory hypotheses that may explain the presence of narrow bandwidth signals in the absence of wide bandwidth power measurements. The West element data indicates a similar shortage of 954 Hz bandwidth power measurement outliers.

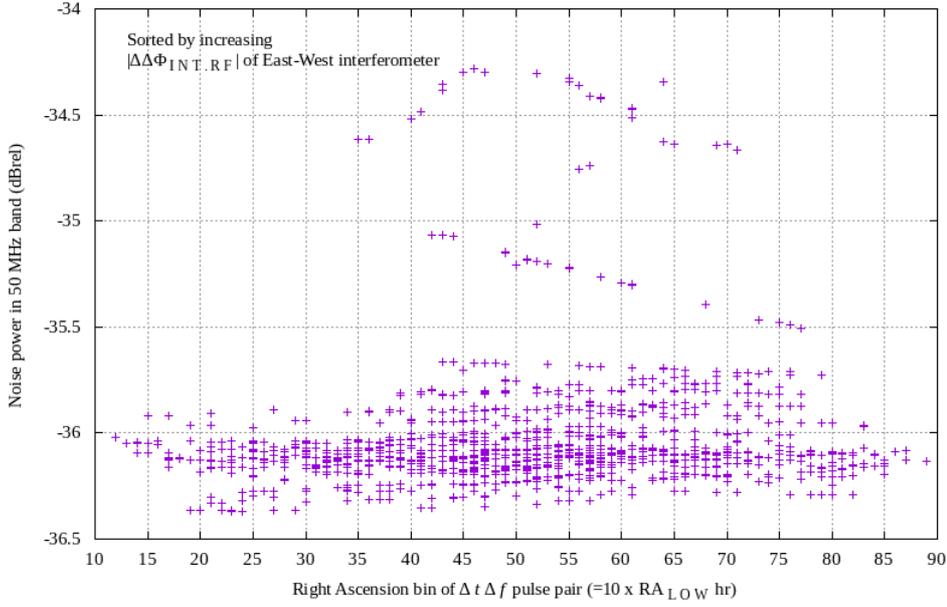

**Figure 6:** A measurement is made of 50 MHz bandwidth integrated East element power, during a 0.27 s FFT time window equal to that of the anomalous $\Delta t=0$ $\Delta f$ pulse pairs. The measurement aids in examining whether or not continuum bandwidth power may explain narrow bandwidth spectral outliers, observed in **Fig. 1-4**. Approximately twenty anomalous *RA* bin 51 and 52 pairs are observed **Fig. 1**, while approximately four anomalous 50 MHz bandwidth *RA* bins 51 and 52 pairs are observed in **Fig. 6**. The measurement of the West element relative average power indicates a similar shortage of anomalous measurements.

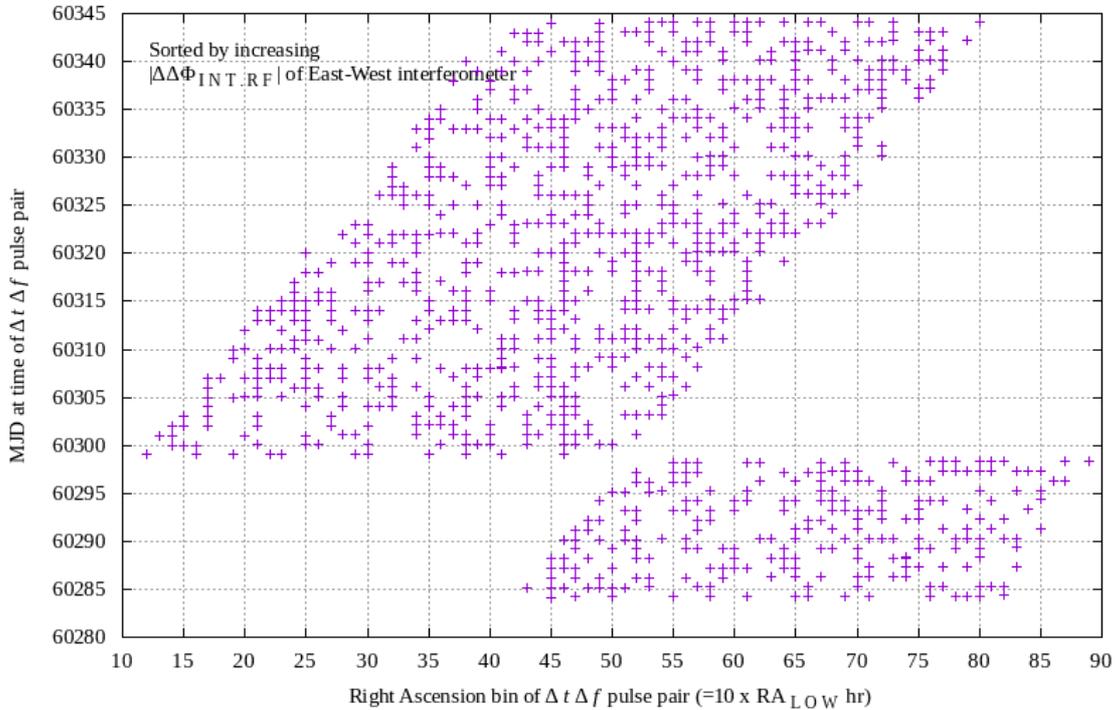

**Figure 7:** The MJD time associated with each polarized pulse pair event is plotted per *RA* bin. Four hour duration files are created, with measured MJD synchronized to UTC using a GPS clock. A single four hour duration file was processed per MJD, while including the 5.25 hr *RA* direction of interest. The number of pulse pair trials vs. *RA* bins have an expected triangular density, assuming a noise cause, which was measured and is shown in **Fig. 9**.

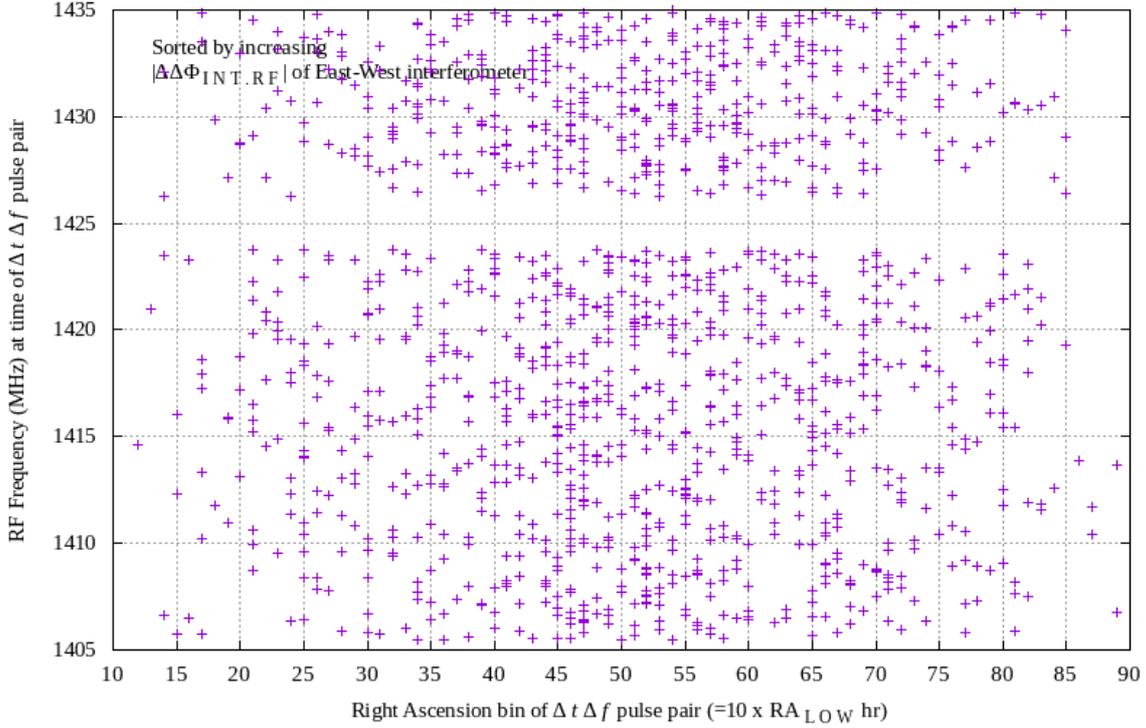

**Figure 8:** The RF frequency of the upper frequency of each of the $\Delta t=0$ $\Delta f$ polarized pulse pairs is measured. Examination of RF frequency distribution is important because RFI is known to often be concentrated in RF frequency. High information capacity communication signals are expected to be approximately uniformly distributed in RF frequency.

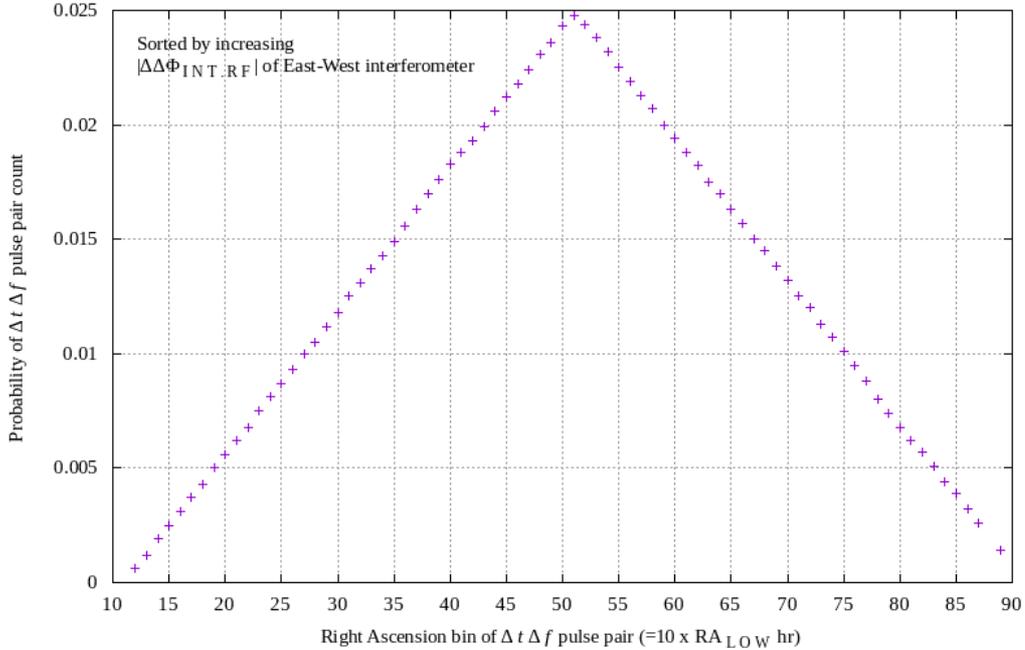

**Figure 9:** The probability of an *RA* bin randomly indicating a pulse pair event, due to noise, is calculated by accumulating the daily time-range contributions of each *RA* bin, then normalizing. The result shows the expected triangular binomial event density, as may be gleaned by observing **Fig. 7**. The probability values in **Fig. 9** are used to determine the two probabilities in the binomial Likelihood calculations per *RA* bin in **Fig. 2**, event *pr.*, and (1 - event *pr.*), i.e. binomial event/non-event probability values.

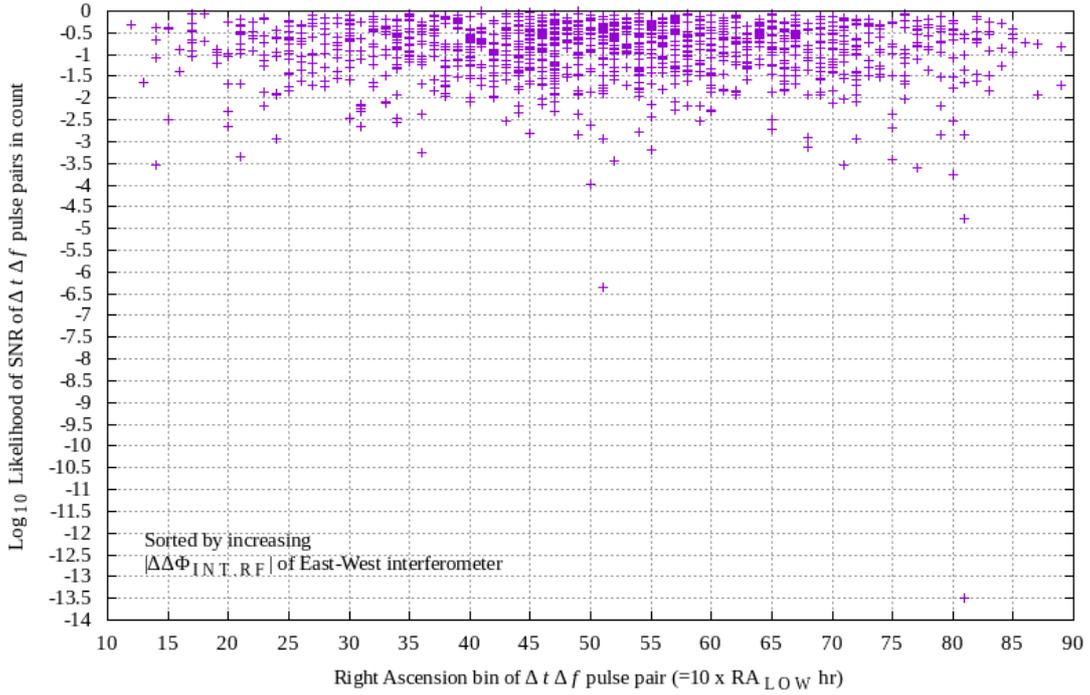

**Figure 10:** *RA* bins 51 and 52 indicate three possibly anomalous SNR measurements, while the majority of measurements in these two *RA* bins does not have many apparent outliers, relative to surrounding *RA* bins.

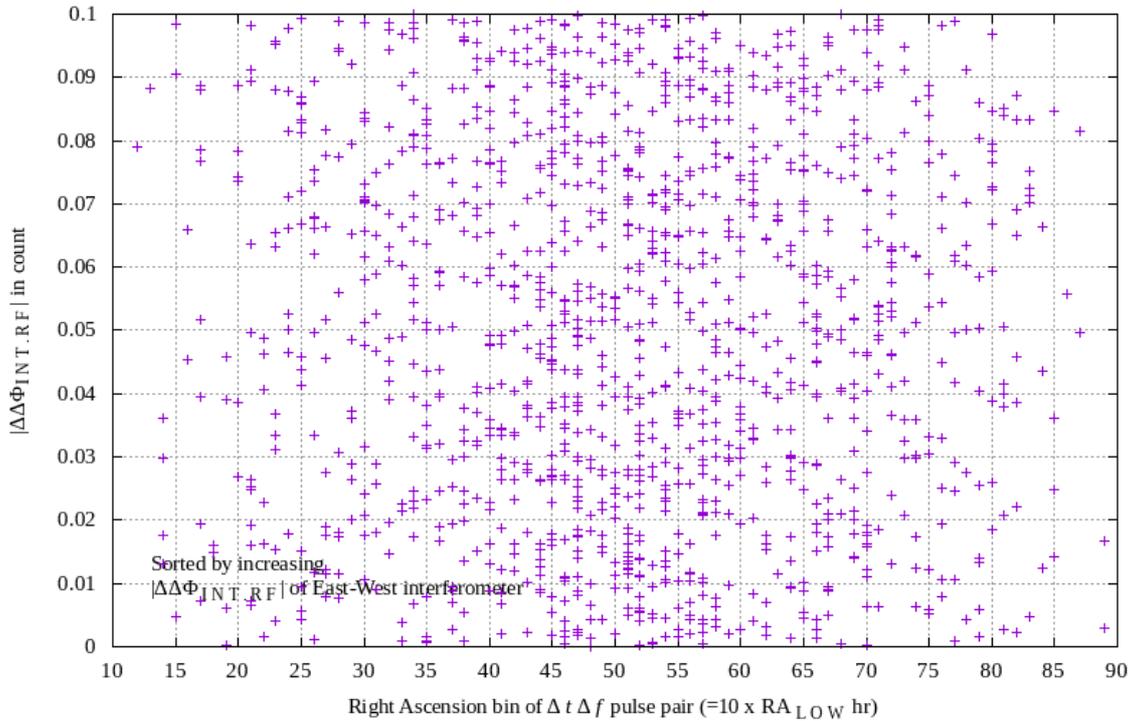

**Figure 11:** $|\Delta\Delta\Phi_{INT.RF}|$, prior to filtering and resulting from an AWGN explanatory model, is estimated to be uniformly-distributed over 0 to $\pi$. Exceptions include pulse pairs having a Ricean amplitude probability distribution, i.e. comprising a sine-wave signal and noise. The statistical power of low-valued $|\Delta\Delta\Phi_{INT.RF}|$ measurements is a topic of ongoing and future work.

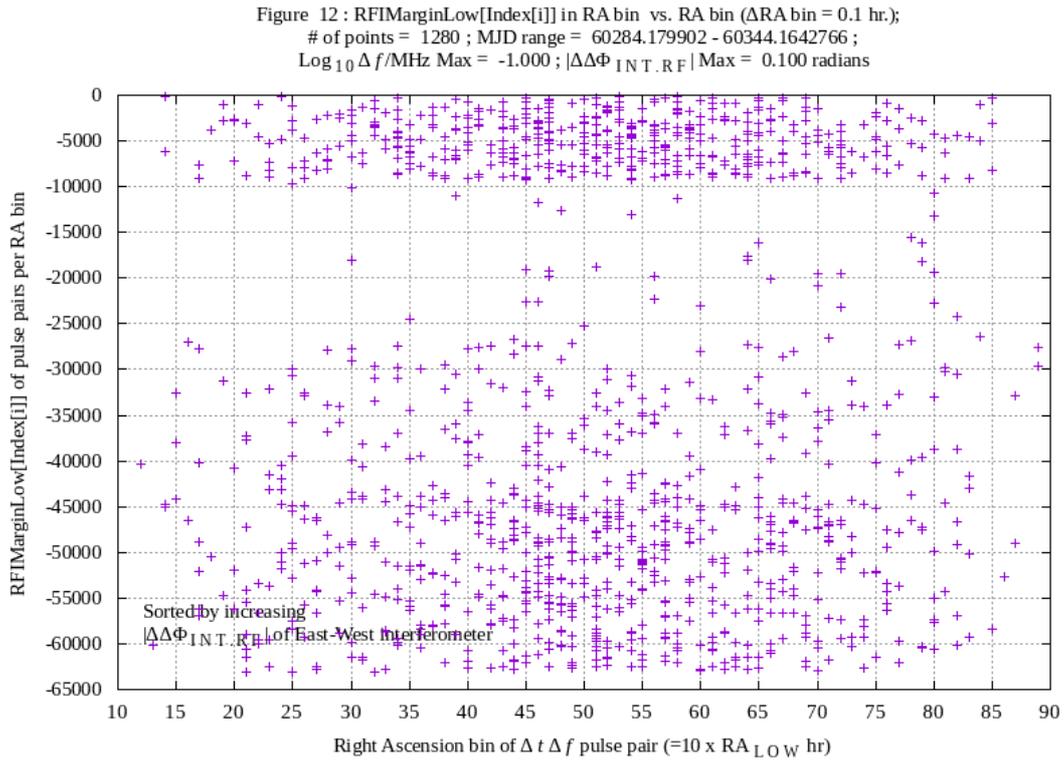

**Figure 12:** The low-sided RFI spectral margin is measured in units of 954 Hz sized segments, to provide potential indications that *RA* bins are corrupted by proximity to nearby excised RFI segments. If excessive numbers of RFI margin measurements have values near zero segment count differences, an RFI cause may be implied.

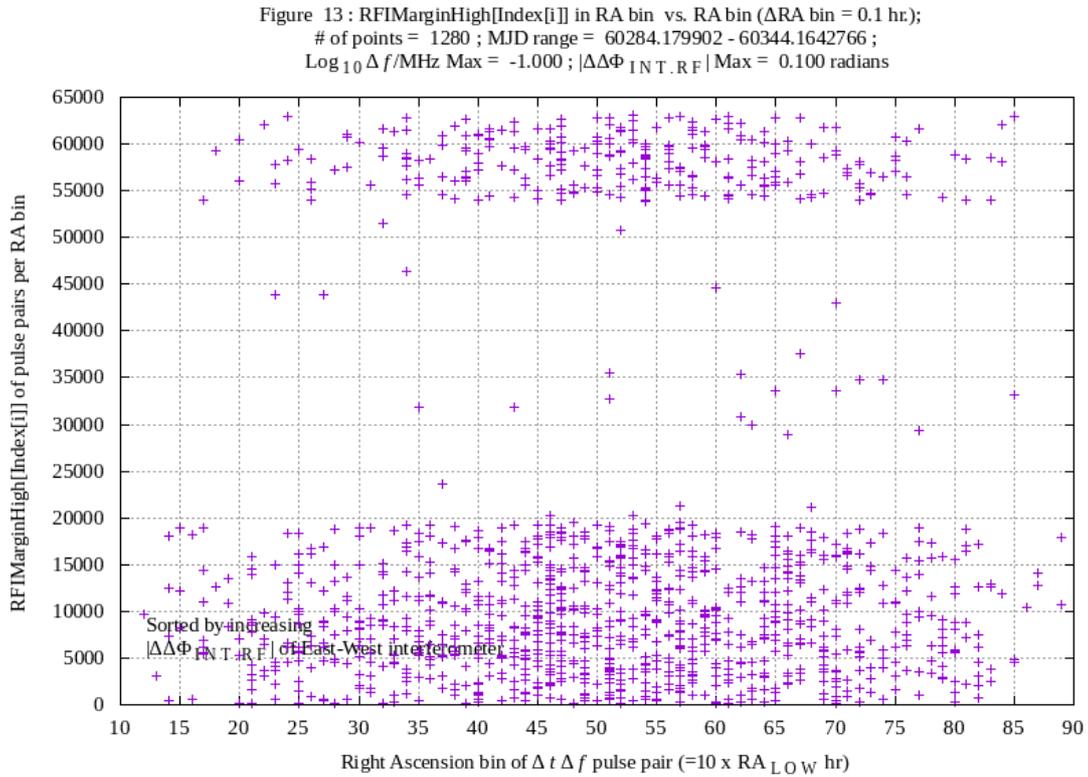

**Figure 13:** In the spectral direction opposite to that shown in **Fig. 12**, RFI margin is examined. These types of RFI margin measurements may be used to help contribute to the falsification of one type of RFI hypotheses, a topic of further work.

## V. Discussion

*The 5.25 hr RA direction indicates anomalies*

The 5.25 ± 0.15 hr *RA*, -7.6° ± 1° *DEC* celestial direction has been associated with a significant number of anomalous pulse pair events in experimental work, since 2018.[5][6][7][8] In the current work, **Fig. 1** presents apparent confirmation that the *RA* bin 52 celestial direction is significant. **Fig. 2** adds statistical power to the *RA* bin 52 importance, because *RA* bin 52 shows a high number of large Cohen's **d** values. The grouping of the points around **d** = 2.5 in **Fig. 2** indicates that the number of anomalous *RA* bin 52 pulse pair events was sustained throughout the 1,280 sorted second-level candidate pulse pair events.

*Choice of sorting parameter*

The choice of the 1,280-sized heap sorting parameter, prior to calculating binomial statistics, may influence the effect size calculation, because sorting parameters that are correlated with a confounding scenario may skew results to the top of the sorted heap. For example, unusually high SNR is often correlated with the presence of RFI. If highest-to-lowest SNR is used to sort the full set of trials in the unsorted heap, RFI events may falsely appear to add significance, due to a potentially high number of RFI events observed at the top of the subsequently sorted heap. In general one should sort using a parameter that is correlated with the experimental characteristic that is being sought. In this experiment, low $\Delta f$ pulse pair sky angle of arrival is sought.

The differential interferometer RF phase $|\Delta\Delta\Phi_{INT.RF}|$ was chosen as the sorting parameter because, among other measurements, $|\Delta\Delta\Phi_{INT.RF}|$ is thought to be the most indicative of pulse pair events having close to the same sky pointing direction, at low $\Delta f$, and in the same FFT time interval, i.e. $\Delta t$=0.

*Associated measurements at $\Delta t = 0$*

Two associated measurements, i.e. the 954 Hz and 50 MHz bandwidth average power measurements, shown in **Figs. 5-6,** do not show strong evidence of apparent response at the *RA* bins that show numerous anomalies, e.g. in **Figs. 1-4,** bins 47 and 52.

*Uncorrelated and correlated noise*

Each interferometer element has an equivalent receiver input noise power that is uncorrelated across elements, as well as a receiver input noise power that is correlated across elements. Correlated noise power across elements is expected to cause an increase in $\Delta t$=0 $\Delta f$ polarized pulse pair counts, due to correlated spectral outliers, while not causing anomalies in low $\Delta f$ measurements, because AWGN is expected to have a Poisson distribution of $\Delta f$ measurements.[5]

Therefore, an increase in 954 Hz bandwidth noise power, due to an astronomical object, for example, may contribute to an increased count of polarized pulse pair candidates. Low $\Delta f$ measurements, on the other hand, allows one to differentiate communication-like pulse pairs from astronomical object caused pulse pairs.

Further work is required to model and quantify correlated and uncorrelated noise measurements.

*Absence of independent corroboration*

To the author's knowledge, independent corroboration of pulsed signals in the *RA, DEC* direction of interest has not been conducted.

## VI. Conclusions

The hypothesis in this work explains an experimental presence of $\Delta t$ $\Delta f$ polarized pulse pairs expected to be due to AWGN.

Difficulties arise while trying to conclude that AWGN explains 5.25 hr *RA* anomalies observed in this experiment, as follows:
(1) the variations in pulse pair counts across *RA*, shown in **Fig. 1**, is inconsistent with an AWGN explanation, and
(2) the statistical power of anomalous concentrations of pulse pair count statistics, shown in **Fig. 2**, is inconsistent with an AWGN explanation.

Consequently, the AWGN hypothesis is thought to be falsified in this experiment.

An important future activity is the development of experimental methods and source models to attempt to falsify alternate and auxiliary hypotheses, e.g. involving natural astronomical objects.

## VII. Further Work

Observations reported here that contribute to falsification of the AWGN hypothesis compel alternate explanations of anomalous signal measurements.

Further work is underway to model, simulate and apply various astronomical object signals to the radio interferometer system. Pulsars, MASERs, Fast Radio Bursts, and continuum object simulated signals may be applied in digital and analog forms, to attempt to falsify auxiliary and alternate explanatory hypotheses.

Further work is underway that repeats long duration interferometer measurements. The observations using the *RA* filter, indicating anomalous celestial directions, is under study, using modified explanatory hypotheses.

Work is underway to add a third interferometer element to the system.

## VIII. Acknowledgements

Many contributions of workers of many organizations, during many years, have made this project highly enjoyable. Special thanks are given to the Green Bank Observatory, National Radio Astronomy Observatory, Deep Space Exploration Society, Society of Amateur Radio Astronomers, SETI Institute, Berkeley SETI Research Center, Breakthrough Listen, Allen Telescope Array, The Penn State Extraterrestrial Intelligence Center, product vendors, and the open source software community. Guidance from family and friends is greatly appreciated.